\documentclass[letter,11pt]{article}
\usepackage{jheppub}

\def\as{\alpha_S}
\def\eps{\varepsilon}
\def\Lb{{\rm L}_\beta}
\def\Lr{{\rm L}_\rho}
\def\b{\beta}

\title{\boldmath NNLO corrections to top pair production at hadron colliders: the quark-gluon reaction}

\author[a]{Micha\l{}  Czakon}
\author[b]{and Alexander Mitov}

\affiliation[a]{Institut f\"ur Theoretische Teilchenphysik und Kosmologie,
RWTH Aachen University, D-52056 Aachen, Germany}
\affiliation[b]{Theory Division, CERN, CH-1211 Geneva 23, Switzerland}

\note{Preprint numbers: CERN-PH-TH/2012-286, TTK-12-45}

\abstract{We compute the next--to--next--to--leading order QCD correction to the total inclusive top pair production cross-section in the reaction $qg \to t\bar t + X$. We find moderate ${\cal O}(1\%)$ correction to central values at both Tevatron and LHC. The scale variation of the cross-section remains unchanged at the Tevatron and is significantly reduced at the LHC. We find that recently introduced approximation based on the high-energy limit of the top pair cross-section significantly deviates from the exact result. The results derived in the present work are included in version 1.4 of the program {\tt Top++}. Work towards computing the reaction $gg\to t\bar t+X$ is ongoing.}

\begin{document} 
\maketitle
\flushbottom

\section{Introduction}

During the last year, the LHC has made major progress in measuring processes with top quarks. As a result, the total production cross-section is now known with few percent accuracy at both 7 and 8 TeV \cite{:2012bt,CMS:note12-007,ATLAS_and_CMS_x-section2012}. The precise measurement of the top pair production cross-section has allowed high-precision extraction of the strong coupling constant \cite{CMS:alpha_s}. At the same time, the mass of the top quark has been measured \cite{CMS:note11-018,ATLAS_and_CMS_mass2012} with precision matching the one from the Tevatron \cite{Aaltonen:2012ra}. 

Such a level of agreement between measurements spanning different colliders, collider energies and final states unambiguously signifies the commencement of the high-precision top quark measurements phase. Given the close relationship between Higgs and top physics \cite{Degrassi:2012ry,Mangano:2012mh}, entering this high-precision phase is particularly significant also in the context of the discovery of a Higgs-like particle \cite{:2012gk,:2012gu} at the LHC.

Equally impressive are the theoretical top physics developments of the recent past. During the last couple of years a number of calculations with NLO accuracy were performed, that accounted for the decay of the top quarks and even off-shell effects  \cite{Biswas:2010sa,Denner:2010jp,Bevilacqua:2010qb,Garzelli:2011vp,Bevilacqua:2011aa,Garzelli:2011is,Melnikov:2011qx,Campbell:2012uf,Denner:2012yc}.  Predictions for the total inclusive cross-section \cite{Ahrens:2010zv,Ahrens:2011mw,Ahrens:2011px,Langenfeld:2009wd,Beneke:2010fm,Kidonakis:2010dk,Kidonakis:2011ca,Beneke:2011mq,Cacciari:2011hy} beyond NLO \cite{Nason:1987xz,Beenakker:1988bq,Czakon:2008ii} were, until recently, exclusively based on using NNLL soft gluon resummation \cite{Beneke:2009rj,Czakon:2009zw,Ahrens:2010zv} and the threshold approximation \cite{Beneke:2009ye} of the partonic cross-section. Very recently also the high-energy limit of the cross-section was incorporated in Ref.~\cite{Moch:2012mk}.

As was demonstrated in Ref.~\cite{Cacciari:2011hy}, see also the discussion in \cite{Kidonakis:2011ca,Beneke:2012wb}, predictions based on soft-gluon resummation alone show only modest improvement over the NLO result, since effects that are subleading in the soft limit can be numerically as significant. 

The first step towards top pair production in NNLO QCD was undertaken in Ref.~\cite{Baernreuther:2012ws}, where the dominant correction from the $q\bar q$ partonic reaction was computed. The remaining, numerically subdominant contribution from this partonic reaction was presented in Ref.~\cite{Czakon:2012zr}, together with the NNLO corrections from the $qq,qq'$ and $q\bar q'$ initiated reactions. 

The results of \cite{Baernreuther:2012ws} demonstrate the importance of a complete NNLO calculation and the role it plays in reducing the theoretical uncertainty. Motivated by this observation, in this work we compute the NNLO correction to the reaction $qg\to t\bar t+X$. Our aim is to verify the effect of this nominally subdominant reaction and check the quality of the approximation \cite{Moch:2012mk} derived from the high-energy limit of the $qg$ partonic cross-section.

The paper is organized as follows: in section~\ref{sec:notation} we introduce our notation. In section~\ref{sec:factorization} we work out the subtraction of the initial state collinear singularities and the evaluation of scale dependent terms. The NNLO parton level result for the reaction $qg\to t\bar t+X$ is presented in section~\ref{sec:results}. In section~\ref{sec:discussion} we discuss the properies of the new NNLO result at parton and hadron levels and compare with existing approximations in the literature. In section~\ref{sec:pheno} we update our ``best" \cite{Cacciari:2011hy} LHC prediction.

\section{Notation}\label{sec:notation}

We follow the notation established in Refs.~\cite{Baernreuther:2012ws,Czakon:2012zr}. At leading power, the total inclusive top pair production cross-section factorizes
\begin{equation}
\sigma_{\rm tot} = \sum_{i,j} \int_0^{\beta_{\rm max}}d\beta\, \Phi_{ij}(\beta,\mu^2)\, \hat\sigma_{ij}(\beta,m^2,\mu^2)  + {\cal O}(\Lambda_{\rm QCD})\, .
\label{eq:sigmatot}
\end{equation}
The indices $i,j$ run over all possible initial state partons; $\beta_{\rm max} \equiv \sqrt{1-4m^2/S}$; $\sqrt{S}$ is the c.m. energy of the hadron collider and $\beta=\sqrt{1-\rho}$, with $\rho\equiv 4m^2/s$, is the relative velocity of the final state top quarks with pole mass $m$ and partonic c.m. energy $\sqrt{s}$. 

The function $\Phi$ in Eq.~(\ref{eq:sigmatot}) is the partonic flux 
\begin{equation}
\Phi_{ij}(\beta,\mu^2) = {2\beta \over 1-\beta^2}~ {\cal L}_{ij}\left({1-\beta_{\rm max}^2\over 1-\beta^2}, \mu^2\right) \, ,
\label{eq:flux}
\end{equation}
expressed through the usual partonic luminosity
\begin{equation}
{\cal L}_{ij}(x,\mu^2) = x \left( f_i\otimes f_j \right) (x,\mu^2) = x \int_0^1 dy \int_0^1 dz \, \delta(x-yz) f_i(y)f_j(z) \, .
\label{eq:Luminosity}
\end{equation}

As usual, $\mu_{R,F}$ are the renormalization and factorization scales. Setting $\mu_F=\mu_R=\mu$, the NNLO partonic cross-section can be expanded through NNLO as
\begin{eqnarray}
\hat\sigma_{ij}\left(\beta,m^2,\mu^2\right) = {\as^2\over m^2}\Bigg\{  \sigma^{(0)}_{ij} + \as \left[ \sigma^{(1)}_{ij} + L\, \sigma^{(1,1)}_{ij} \right] + \as^2\left[ \sigma^{(2)}_{ij} + L\, \sigma^{(2,1)}_{ij} + L^2 \sigma^{(2,2)}_{ij} \right] \Bigg\} \, .
\label{eq:sigmapart}
\end{eqnarray}
In the above equation $L = \ln\left(\mu^2/m^2\right)$, $\as$ is the ${\overline {\rm MS}}$ coupling renormalized with $N_L=5$ active flavors at scale $\mu^2$ and $\sigma^{(n(,m))}_{ij}$ are functions only of $\beta$.

All partonic cross-sections are known exactly through NLO \cite{Nason:1987xz,Beenakker:1988bq,Czakon:2008ii}. The scaling functions $\sigma^{(2,1)}_{ij}$ and $\sigma^{(2,2)}_{ij}$ can be computed from $\sigma^{(1)}_{ij}$, see section \ref{sec:factorization}. The dependence on $\mu_R\neq \mu_F$ can be trivially restored in Eq.~(\ref{eq:sigmapart}) by re-expressing $\as(\mu_F)$ in powers of $\as(\mu_R)$; see for example Ref.~\cite{Langenfeld:2009wd}. 

The reactions $ij\to t\bar t + X$ were computed for $i,j = (q\bar q,qq,qq',q\bar q')$ through NNLO in Refs.~\cite{Baernreuther:2012ws,Czakon:2012zr}. In this paper we compute the NNLO correction to the reaction $qg\to t\bar t + X$. The only currently unknown contribution to $t\bar t$ production at NNLO is the $gg$ initiated reaction, which will be the subject of a future publication.

\section{Collinear factorization and scale dependence}\label{sec:factorization}

We follow the setup and notation described in Ref.~\cite{Czakon:2012zr} and denote the collinearly unrenormalized partonic cross-sections as $\tilde \sigma_{ij}^{(n)}(\eps,\rho)$. Then, introducing the functions $\tilde s^{(n)}_{ij}$ and $s^{(n)}_{ij}$ defined as $\tilde s_{ij}^{(n)}(\eps,\rho) \equiv \tilde\sigma^{(n)}_{ij}(\eps,\rho)/\rho$ and $s^{(n)}_{ij}(\rho) \equiv \sigma^{(n)}_{ij}(\rho)/\rho$, the $\overline{\rm MS}$--subtracted $qg$-initiated cross-section $s^{(n)}_{qg}$ reads through NNLO:
\begin{eqnarray}
s^{(1)}_{qg} &=& \tilde s^{(1)}_{qg} + {1\over \epsilon} \left({1\over 2\pi}\right) \Bigg\{ \tilde s^{(0)}_{q \bar q} \otimes P^{(0)}_{qg}  + \tilde s^{(0)}_{gg} \otimes P^{(0)}_{gq}\Bigg\}  \, , \label{eq:shat1}\\ 
&& \nonumber\\
s^{(2)}_{qg} &=& \tilde s^{(2)}_{qg} +  \left({1\over 2\pi}\right)^2 \Bigg\{ -{\beta_0\over 2\epsilon^2}\left[ \tilde s^{(0)}_{gg} \otimes P^{(0)}_{gq}+ \tilde s^{(0)}_{q \bar q} \otimes P^{(0)}_{qg} \right]  + {1\over 2\epsilon}\left[  \tilde s^{(0)}_{gg} \otimes P^{(1)}_{gq}+ \tilde s^{(0)}_{q \bar q} \otimes P^{(1)}_{qg} \right] \nonumber\\
&& + {1\over 2\epsilon^2}\left[ 3 \tilde s^{(0)}_{gg} \otimes P^{(0)}_{gg}\otimes P^{(0)}_{gq} + \tilde s^{(0)}_{gg} \otimes P^{(0)}_{gq}\otimes P^{(0)}_{qq} +3\tilde s^{(0)}_{q \bar q} \otimes P^{(0)}_{qq}\otimes P^{(0)}_{qg} + \tilde s^{(0)}_{q \bar q} \otimes P^{(0)}_{qg}\otimes P^{(0)}_{gg} \right] \Bigg\} \nonumber\\
&&+  {1\over \epsilon} \left({1\over 2\pi}\right) \Bigg\{ \tilde s^{(1)}_{q \bar q} \otimes P^{(0)}_{qg}  + \tilde s^{(1)}_{qg} \otimes P^{(0)}_{gg} + \tilde s^{(1)}_{qg} \otimes P^{(0)}_{qq} + \tilde s^{(1)}_{gg} \otimes P^{(0)}_{gq}\Bigg\} \, ,
\label{eq:shat2}
\end{eqnarray}
with $\beta_0=11C_A/6-N_L/3$.

The integral convolutions in Eq.~(\ref{eq:shat2}) are performed numerically, over a set of 80 points in the interval $\beta \in (0,1)$.  The only non-trivial step in this evaluation is the derivation of the partonic cross-section $\tilde s^{(1)}_{gg}$ through order ${\cal O}(\epsilon)$.  To derive it, we follow the approach of Ref.~\cite{Czakon:2008ii} which allows one to derive analytical results for the required partonic cross-sections. The order ${\cal O}(\epsilon)$ terms of $\tilde s^{(1)}_{q\bar q}$ and $\tilde s^{(1)}_{qg}$ can be easily computed this way and expressed in terms of standard harmonic polylogarithms (HPL) \cite{Remiddi:1999ew}. As can be anticipated from the findings of Ref.~\cite{Czakon:2008ii}, however, the calculation of $\tilde s^{(1)}_{gg}$ through order ${\cal O}(\epsilon)$ introduces a number of new functions that go beyond the class of HPL's. In particular, some functions are represented as two dimensional integrals. From a numerical point of view, this is problematic since it significantly reduces the speed of the numerical integrations  in Eq.~(\ref{eq:shat2}). To deal with the loss of speed, we have resorted to  interpolation techniques, which limits the appeal (and usefulness) of an intrinsically analytic approach. Based on our experience, we conclude that such an approach for the computation of the collinear factorization contributions is suboptimal.

The evaluation of the scale dependent functions $\sigma^{(2,1)}_{qg}$ and $\sigma^{(2,2)}_{qg}$ is rather straightforward, see \cite{Czakon:2012zr} for details. In terms of the functions $s^{(n(,m))}_{ij}(\rho) \equiv \sigma^{(n(,m))}_{ij}(\rho)/\rho$ we get:
\begin{eqnarray}
s^{(2,2)}_{qg} &=& {1\over 2 (2\pi)^2} \left[ 
-5 \beta_0\left(s^{(0)}_{gg}\otimes P^{(0)}_{gq} + s^{(0)}_{q\bar q}\otimes P^{(0)}_{qg}\right) + 
3 s^{(0)}_{gg}\otimes P^{(0)}_{gq}\otimes P^{(0)}_{gg} \right.\nonumber\\
&+&\left. s^{(0)}_{gg}\otimes P^{(0)}_{gq}\otimes P^{(0)}_{qq}+3 s^{(0)}_{q\bar q}\otimes P^{(0)}_{qq}\otimes P^{(0)}_{qg} + s^{(0)}_{q\bar q}\otimes P^{(0)}_{qg}\otimes P^{(0)}_{gg}\right] \, ,\nonumber\\
s^{(2,1)}_{qg} &=& -{1\over (2\pi)^2} \left[ s^{(0)}_{gg}\otimes P^{(1)}_{gq} + s^{(0)}_{q\bar q}\otimes P^{(1)}_{qg}\right] \nonumber\\
&+& {1\over 2\pi} \left[3\beta_0 s^{(1)}_{qg} - s^{(1)}_{gg}\otimes P^{(0)}_{gq} - s^{(1)}_{qg}\otimes P^{(0)}_{gg} - s^{(1)}_{qg}\otimes P^{(0)}_{qq} - s^{(1)}_{q\bar q}\otimes P^{(0)}_{qg}\right]\, .
\label{scales-qtildeq}
\end{eqnarray}

Eq.~(\ref{scales-qtildeq}) agrees with Ref.~\cite{Langenfeld:2009wd}. The convolutions appearing in Eq.~(\ref{scales-qtildeq}) are computed numerically. We have checked that the fits implemented in the program Hathor \cite{Aliev:2010zk} agree with our own numerical calculation of Eq.~(\ref{scales-qtildeq}) to a very high precision. Given this level of agreement, instead of producing new fits, we have implemented the analytical fits for $\sigma^{(2,1)}_{qg}$ and $\sigma^{(2,2)}_{qg}$ from Ref.~\cite{Aliev:2010zk} in our program {\tt Top++} (ver 1.4) \cite{Czakon:2011xx}.

\section{Parton level results}\label{sec:results}

For the calculation of the collinearly unrenormalized partonic cross-section $\tilde \sigma^{(2)}_{qg}$ we follow the approach already used in Refs.~\cite{Baernreuther:2012ws,Czakon:2012zr}. The correction due to double real radiation is computed following Refs.~\cite{Czakon:2010td,Czakon:2011ve}.
\footnote{Methods for computing the double real radiation for this process have also been developed in Refs.~\cite{Abelof:2011ap,Abelof:2012rv,Abelof:2012he}.}
For the real-virtual correction we use the counter-terms from Refs.~\cite{Bern:1999ry,Catani:2000pi,Catani:2000ef,Bierenbaum:2011gg}. For the evaluation of the required one-loop five-point amplitude we use a code from the calculation of $pp\to t\bar t + {\rm jet}$ at NLO \cite{Dittmaier:2007wz}. 

As in Refs.~\cite{Baernreuther:2012ws,Czakon:2012zr}, Eq.~(\ref{eq:sigmapart}) is derived in a renormalization scheme where the number of active flavors $N_f$ equals the number of light flavors, i.e. $N_f=N_L=5$. From a practical point of view, the calculation is performed in three steps. In the first step all calculations, including UV renormalization, are performed in a standard way by working in conventional dimensional regularization (CDR) and considering all fermions as active flavors, i.e. $N_f=N_L+1$. The renormalization procedure, including the relevant renormalization constants, has been described, for example, in \cite{Czakon:2007wk,Czakon:2007ej}. In the second step the heavy flavor is decoupled.  The decoupling procedure is applied in $d=4-2\epsilon$ dimensions, as appropriate, to each of the principal contributions to the cross-section: double-real, real-virtual and if present, one- and two-loop virtual amplitudes. The decoupling constant can be found, for example, in Ref.~\cite{Czakon:2007wk}. The third and final step consists of the collinear subtraction described around Eq.~(\ref{scales-qtildeq}), which is performed with all cross-sections (obtained in step two), splitting functions and $\beta$-function coefficients evaluated consistently in a scheme with $N_f=N_L=5$ active flavors.

The result for the NNLO correction to the reaction $qg\to t\bar t +X$ reads:
\begin{eqnarray}
\sigma^{(2)}_{qg}(\beta) = F_0(\beta) +F_1(\beta) N_L \, .
\label{eq:sigma2qg}
\end{eqnarray}
The full dependence on the number of light flavors $N_L$ in Eq.~(\ref{eq:sigma2qg}) is made explicit. The functions $F_{0,1}$ read:
\begin{eqnarray}
F_1 &=&  0.363838 \b^2 - 1.44391 \b^3 + 1.1146 \b^7 - 0.309165 \b^3 \Lb + 0.990057 \b^4 \Lb^2 + 0.362183 \rho^2 \Lr \nonumber\\
&+& \left(0.194867 \rho + 1.57274 \rho^2\right) \Lr^2 + 0.0401411 \rho \Lr^3 \, , \label{eq:F1}\\ 
F_0 &=& 28.0998 \b^2 + 24.1753 \b^3 - 12.3211 \b^5 - 49.909 \b^7 + 11.7853 \b^3 \Lb + 28.6697 \b^6 \Lb^2 \nonumber\\
&+& \left(-1.68957 + 30.6335 \rho^2\right) \Lr + \left(-9.80339 \rho - 76.7407 \rho^2\right) \Lr^2 - 3.82993 \rho \Lr^3 \, , \label{eq:F0} 
\end{eqnarray}
where $\Lr \equiv \ln(\rho)$, $\Lb \equiv \ln(\beta)$ and we recall that $\rho = 1-\b^2$. 

\begin{figure}[tbp]
\centering
\includegraphics[width=.95\textwidth]{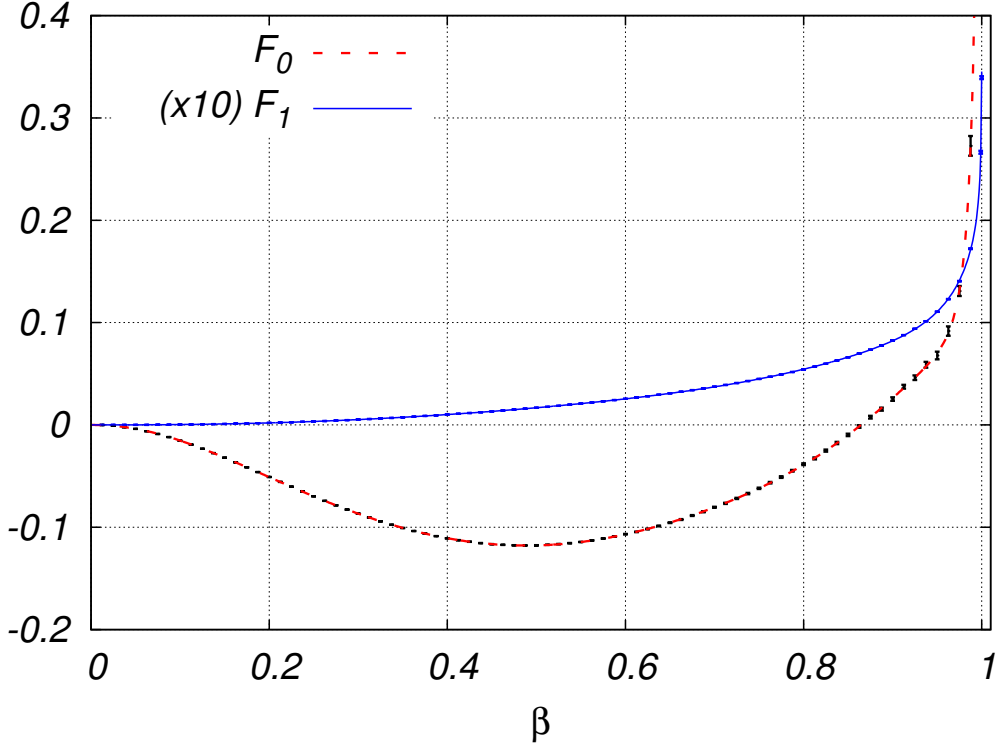}
\caption{\label{fig:data-and-fits} The functions $F_0(\b)$ and $F_1(\b)$ (the latter multiplied by a factor of 10 for better visibility) as defined in Eq.~(\ref{eq:sigma2qg}). Shown are the fits for $F_0(\b)$ (dashed red) (\ref{eq:F0}), $F_1(\b)$ (solid blue) (\ref{eq:F1}) and the discrete computed values, including their numerical errors.}
\end{figure}

The functions $F_{0,1}$ in Eqs.~(\ref{eq:F1},\ref{eq:F0}) are fits to the numerically computed partonic cross-section. The calculation of  the function $F_{0}$ is done in 80 points in the interval $\beta \in (0,1)$. The highest computed point is $\b_{80} = 0.999$. The fit and the computed points, including their numerical errors, are shown in fig.~\ref{fig:data-and-fits}. 
Except for the very last point $\b_{80}$, the quality of the calculation is high, sub-1\%. The quality of the fit is also good; it fits the computed points within the numerical uncertainties for large and moderate $\b$. For smaller values of $\b$ the quality of the fit is not as high, yet the relative deviation of the fit from the computed mid-points is better than 1\%. Only in the region of very small $\b$ the relative distance between the fit and computed central values surpasses 1\% but in that range the absolute size of the result is negligible which makes such  deviation phenomenologically irrelevant. 

The calculation of the function $F_{1}$ is done in 81 points in the interval $\beta \in (0,1)$. In addition to the 80 points used in the computation of $F_{0}$, we have added the extra point $\b=0.99999$. The fit and all computed points, including their numerical errors, are shown in fig.~\ref{fig:data-and-fits}. The quality of the fit is high, sub-1\%, for $\b\gtrsim 0.1$. In the region of smaller $\b$ the quality of the fit deteriorates, but the absolute difference between the fit and computed points is extremely small and also phenomenologically irrelevant. 

The most prominent feature of the partonic cross-section $\sigma_{qg}^{(2)}$ is its high-energy behavior \cite{Nason:1987xz,Catani:1990xk,Collins:1991ty,Catani:1990eg,Catani:1993ww,Catani:1994sq}:
\begin{equation}
\sigma^{(2)}_{qg\to t\bar t + X} \Big\vert_{\rho \to 0} \approx c_1\ln(\rho) + c_0 +{\cal O}(\rho) \, .
\label{eq:high-energy-limit}
\end{equation}
The constant $c_1$ has been predicted exactly in Ref.~\cite{Ball:2001pq}, with $N_L$-independent numerical value
\begin{equation}
c_1 = -1.689571450230512 \, .
\label{eq:c1}
\end{equation}

To improve the high-energy endpoint behavior of the fits (\ref{eq:F1},\ref{eq:F0}), we have imposed on them the exact $\sim\ln(\rho)$ behavior from (\ref{eq:high-energy-limit}). Then, from the fits (\ref{eq:F1},\ref{eq:F0}), we derive an estimate of the constant $c_0$ appearing in Eq.~(\ref{eq:high-energy-limit}). We get the value:
\begin{equation}
c_0 = -9.96 + 0.0345 N_L \, .
\label{eq:c0}
\end{equation}

Setting $N_L=5$ we find that Eq.~(\ref{eq:c0}) agrees 
\footnote{We note that the prediction for the constant $c_0$ derived in Ref.~\cite{Moch:2012mk} contains no explicit $N_L$ dependence.}
with the numerical estimate of $c_0$ derived  in Ref.~\cite{Moch:2012mk} with the help of completely independent methods.

The numerical error on the proportional to $N_L$  term in Eq.~(\ref{eq:c0}) is likely rather small, thanks to our ability to extend the calculation of $F_1$ to $\b$ as high as $\b=0.99999$ and to the fact that the function $F_1$ behaves $\sim const$ at large $\b$. 

On the other hand, estimating the error on the $N_L$-independent part of Eq.~(\ref{eq:c0}) is much harder. The reason for this is that the region below $\b_{80}=0.999$ (which is the highest computed point for $F_0$) is still not close enough to the high-energy endpoint to be dominated by the high-energy expansion (\ref{eq:high-energy-limit}). Going beyond the highest computed point $\b_{80} = 0.999$ is currently unfeasible since the computational cost for a single point, located well above the point $\b_{80}$,  would be comparable to the computational cost for all 80 calculated points.

Combining the above observations with the fact that the numerical error in this last computed point is larger, exceeding 1\%, we conclude that the error on the $N_L$-independent part of $c_0$ could possibly be as large as few tens of percent.

\section{Discussion}\label{sec:discussion}

\subsection{Properties of the parton level result}

The most striking feature of the ${\cal O}(\as^4)$ correction to $\hat\sigma_{qg}$ is the similarity of its shape and size to the long known ${\cal O}(\as^3)$ correction.
\footnote{For short, in the rest of this section we refer to these two corrections as NNLO and NLO, respectively.}
\begin{figure}[tbp]
\centering
\includegraphics[width=.95\textwidth]{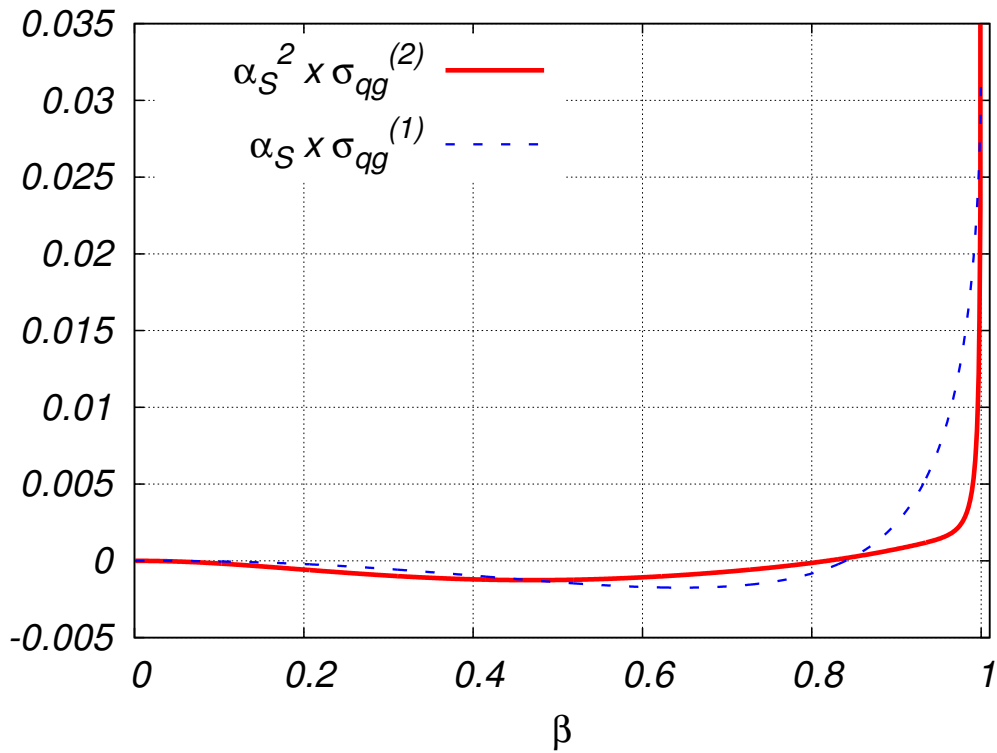}
\caption{\label{fig:gq-NLO-NNLO} Comparison of the NLO and NNLO corrections to the partonic cross-section, including the relative power of $\as(m_t)$ as in Eq.~(\ref{eq:sigmapart}): $\as\sigma_{qg}^{(1)}$ (dashed blue) and $\as^2\sigma_{qg}^{(2)}$ (solid red).}
\end{figure}
In fig.~\ref{fig:gq-NLO-NNLO} we compare the two, including the appropriate relative powers of the strong coupling $\as(m_t) \approx 0.1068$, see Eq.~(\ref{eq:sigmapart}). We observe that the main difference between the two curves is in their high-energy behavior, which is more singular in the case of $\sigma_{qg}^{(2)}$. The similarity in size and shape between the two consecutive perturbative corrections indicates that large perturbative NNLO corrections can be expected. Indeed, if it was not for the suppression due to the additional power of $\as$, the NNLO correction could have been even more sizable, a feature that might be relevant for the description of lighter fermion pair production, like bottom quarks. 

\begin{figure}[tbp]
\centering
\hspace{-.2cm} 
\includegraphics[width=7.9cm]{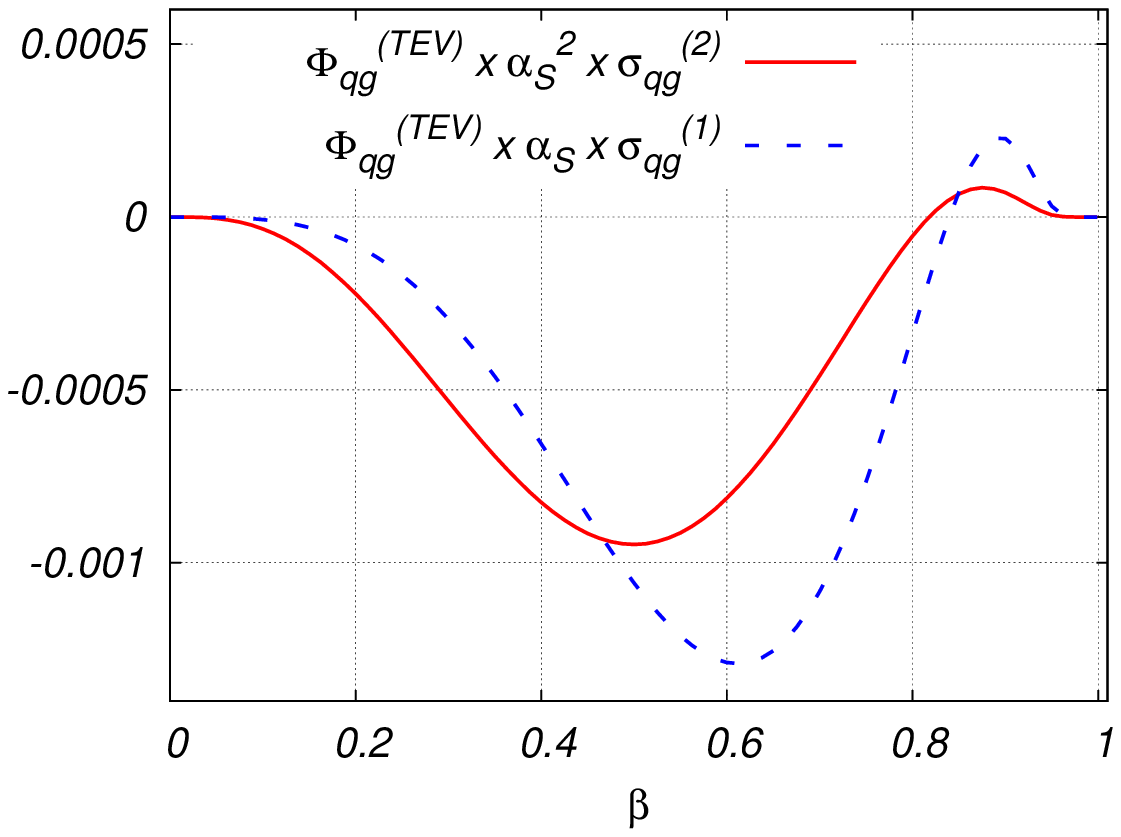}
\hspace{-.4cm} 
\includegraphics[width=7.9cm]{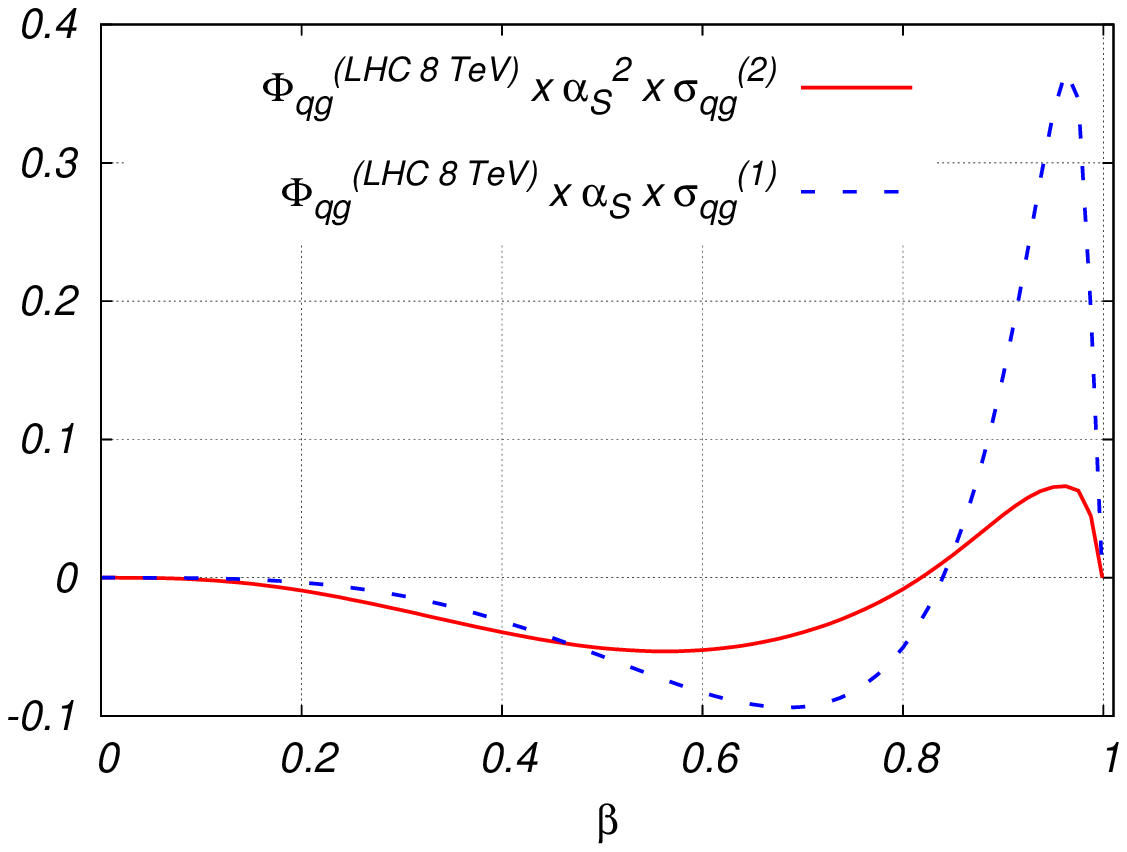}
\caption{\label{fig:fluxes} NLO and NNLO corrections to the partonic cross-section (as in fig.~\ref{fig:gq-NLO-NNLO}) times the flux at Tevatron (left) and LHC 8 TeV (right), see Eq.~(\ref{eq:sigmatot}).}
\end{figure}

To better assess the phenomenological significance of the similarities and differences between the NLO and NNLO corrections, in fig.~\ref{fig:fluxes} we plot their product with the partonic fluxes for the Tevatron and LHC 8 TeV, see Eq.~(\ref{eq:sigmatot}) for precise definition. 
In all cases we use MSTW2008nnlo68cl pdf set~\cite{Martin:2009iq}. The relative powers of $\as(m_t)$ are also included. 

We observe that the similarity in shape and size between the two corrections is preserved at the Tevatron. Therefore, one can anticipate NNLO contribution to the hadron-level cross-section $\sigma_{\rm tot}$ that is similar in size to the NLO one. On the other hand, at the LHC 8 TeV, we observe a dramatic difference between the shapes and sizes of the NLO and NNLO corrections. Clearly, at this particular collider energy, one can anticipate very strong cancellation between the positive and negative nodes of the NLO result, while the NNLO one stays mostly negative. We also note that the high-energy rise of the NNLO correction is completely screened by the flux which vanishes in the same limit. 

To better quantify the differences between the NLO and NNLO corrections, in table~\ref{tab:table} we present their separate contributions to $\sigma_{\rm tot}$.
\begin{table}[ht]
\begin{center}
\begin{tabular}{| c | c | c | c | c | c |}
\hline
&& {\rm Tevatron}  & {\rm LHC}~7~{\rm TeV} &  {\rm LHC}~8~{\rm TeV} & {\rm LHC}~14~{\rm TeV} \\
\hline 
$I_1$ & {\rm Due\ to}~$\sigma_{qg}^{(1)}$~[pb] & -0.068 & -0.88 & -0.48 & 9.01 \\ 
\hline 
$I_2$ & {\rm Due\ to}~$\sigma_{qg}^{(2)}$~[pb] & -0.057 & -1.82 & -2.25 & -4.07 \\ 
\hline
$I_3$ & $\sigma_{qg}^{(2)}({\rm Hathor}; (A+B)/2)$~[pb] & 0.040 & 5.78 & 8.11 & 27.36 \\ 
\hline
$I_4$ & $(I_3-I_2)/\sigma_{\rm tot}$~[\%] & 1.4 & 4.9 & 4.7 & 3.7 \\ 
\hline
\end{tabular}
\caption{\label{tab:table} Central values for the contributions of $\sigma_{qg}^{(1)}$ and $\sigma_{qg}^{(2)}$ to $\sigma_{\rm tot}$ for the Tevatron and LHC 7,8 and 14 TeV. Also shown is the corresponding contribution from the program Hathor based on the approximation to $\sigma_{qg}^{(2)}$ of Ref.~\cite{Moch:2012mk}. Line 4 shows the difference between the exact result (this paper) and the approximation from Ref.~\cite{Moch:2012mk}, relative to $\sigma_{\rm tot}$.}
\end{center}
\end{table}
The results on lines 1 and 2 as well as $\sigma_{\rm tot}$ on line 4 are computed with version 1.4 of the program {\tt Top++} \cite{Czakon:2011xx} with default settings, $m_t=173.3~{\rm GeV}$, central scales and MSTW2008nnlo68cl pdf set~\cite{Martin:2009iq}. The numbers in line 3 are computed with the program Hathor \cite{Aliev:2010zk} using the same parameters as above. We comment on the results obtained with Hathor in section~\ref{sec:comparison}.

From the first two rows of table~\ref{tab:table} we conclude that the absolute size of the NNLO correction can be significantly larger than the NLO one depending on the c.m. energy of the collider. Comparing to fig.~\ref{fig:gq-NLO-NNLO}, however, we note that the large differences between the ${\cal O}(\as^3)$ and ${\cal O}(\as^4)$ corrections seen in table~\ref{tab:table} do not necessarily indicate a breakdown of the perturbative expansion, since they result from large accidental cancellations that strongly depend on the collider energy.

\begin{figure}[tbp]
\centering
\includegraphics[width=.95\textwidth]{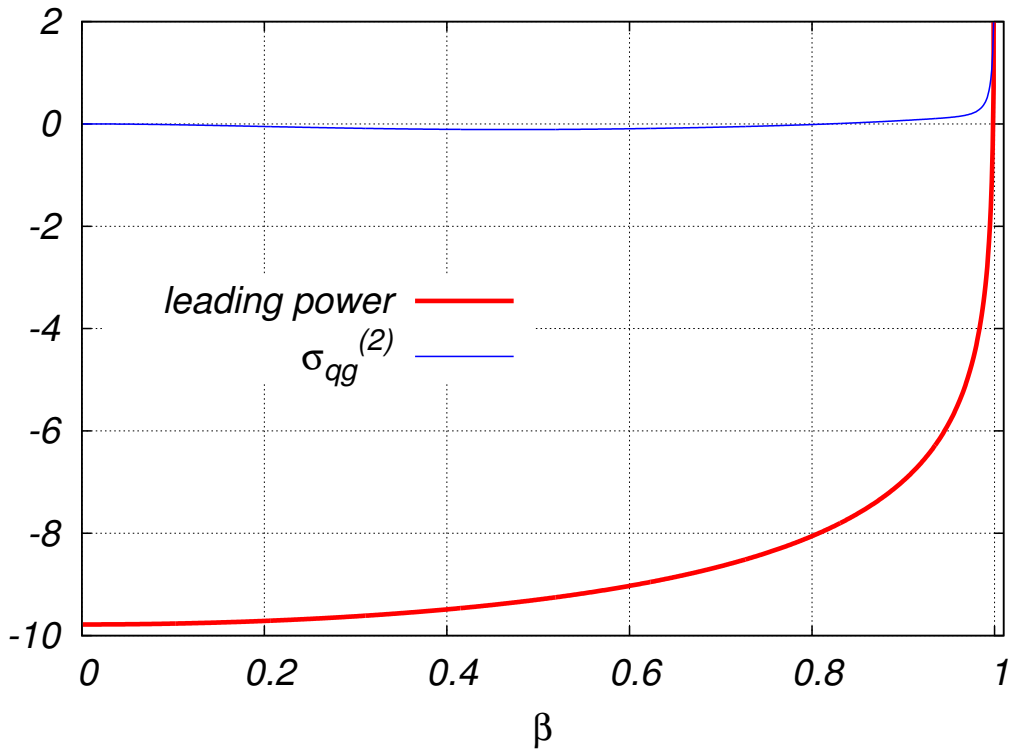}
\caption{\label{comparisons-full_HighEnergy} Comparison of the exact result (\ref{eq:sigma2qg}) for $\sigma_{qg}^{(2)}$ (thin blue line) and the leading power of its high-energy expansion (\ref{eq:high-energy-limit}) (thick red line).}
\end{figure}

Next, in fig.~\ref{comparisons-full_HighEnergy}, we compare the exact result for $\sigma_{qg}^{(2)}$ with its leading-power high-energy approximation (\ref{eq:high-energy-limit}). In complete analogy with the case of fermion pair initiated top pair production \cite{Czakon:2012zr} we observe that the high-energy approximation is justified only very close to the high-energy endpoint and is a poor approximation to the exact result outside of this narrow range.

\subsection{Properties of the hadron level result}\label{sec:had-lelel-prop}

To better judge the effect from the inclusion of the NNLO correction to the $qg$ reaction, in table~\ref{tab:pheno} we give the central values for our best prediction for the Tevatron and LHC 8 TeV in the following cases: 
\begin{itemize}
\item With, or without, soft gluon resummation: the $gg$ reaction is included in NLO+NNLL or in approximate NNLO (defined as in Refs.~\cite{Baernreuther:2012ws,Cacciari:2011hy}), while all other reactions are included in NNLO+NNLL or NNLO.
\item With, or without, the NNLO corrections to $qg\to t\bar t+X$. 
\end{itemize}
Besides the central values, in table~\ref{tab:pheno} we also show the average scale uncertainty, defined as $({\rm scale}_+ + {\rm scale}_-)/2$. As elsewhere in this article we use MSTW2008nnlo68cl pdf set~\cite{Martin:2009iq}, $m_t=173.3$ GeV and scale and pdf variations are performed as described in Ref.~\cite{Cacciari:2011hy}.

\begin{table}[h]
\begin{center}
\begin{tabular}{| l | c | c || c | c |}
\hline
& Resummed;      & Resummed; &  Fixed Order;    & Fixed Order; \\
& $qg$-included & no $qg$      & $qg$-included & no $qg$        \\
\hline 
Tevatron: central [pb] & 7.010 & 7.067 & 6.949 & 7.006 \\ 
\hline 
Tevatron: aver. scale var. [\%]& $\pm$2.6 & $\pm$2.7 & $\pm$4.4 & $\pm$3.7 \\ 
\hline\hline 
LHC 8: central [pb] & 220.4 & 222.7 & 218.5 & 220.8 \\
\hline 
LHC 8: aver. scale var. [\%]& $\pm$5.3 & $\pm$7.3 & $\pm$4.5 & $\pm$6.5 \\ 
\hline
\end{tabular}
\caption{\label{tab:pheno} Central values and average scale variations, the latter defined as $({\rm scale}_+ + {\rm scale}_-)/2$, of our ``best" prediction for the Tevatron and LHC 8 TeV.  Numbers are given for the following four cases: with/without soft gluon resummation and with/without including $\sigma_{qg}^{(2)}$. The values of the various parameters used in the calculation are specified in the text. }
\end{center}
\end{table}

We observe that the effect on the central value from the inclusion of the NNLO $qg$ correction is moderate, and brings down the central value by about $0.8\%$ at the Tevatron and by about $1\%$ at the LHC. Such a shift is perfectly consistent with our estimate \cite{Baernreuther:2012ws} of the theoretical uncertainty at the Tevatron. We also note that, as might be anticipated, the size of the shift in the central value is the same independently of the inclusion (or not) of soft gluon resummation. 

The effects of the NNLO $qg$ correction on the size of the scale variation is more consequential. At the Tevatron, the uncertainty in the pure fixed order prediction increases by about $\pm 0.7\%$, while the uncertainty in the soft gluon resummed result is unaffected by the inclusion of the NNLO $qg$ correction. This is consistent with the expectation that the dominant source of uncertainty at the Tevatron is already accounted for. The effect on the fixed order prediction is also at a level similar to the anticipated \cite{Baernreuther:2012ws} NNLO correction in $gg\to t\bar t+X$. 

At the LHC, on the other hand, we notice a dramatic $\pm2\%$ decrease in scale uncertainty both with and without including soft gluon resummation. This is a significant improvement in the precision of the theoretical prediction at the LHC. Despite this improvement, however, it is clear that the unknown genuinely NNLO correction in the $gg$ reaction still dominates the uncertainty at the LHC. This is evident, for example, from the fact that the scale variation of the resummed result is larger than that of the fixed order result (which we take as a more conservative estimate of the theoretical uncertainty \cite{Cacciari:2011hy}). 

Before closing this section we address the question of how the uncertainty in the derived by us constant $c_0$ (\ref{eq:c0}) propagates into phenomenological predictions. As we argued in Ref.~\cite{Baernreuther:2012ws}, the most natural way to address this question is to consider the ratio:
\begin{equation}
R_{qg}(\b_{80}) =  {\Sigma_{qg}(\b_{80}) \over \Sigma_{qg}(0)} \, ,
\label{eq:R}
\end{equation}
where:
\begin{equation}
\Sigma_{qg}(\b_{80}) = \int_{\b_{80}}^{\beta_{\rm max}}d\beta\, \Phi_{qg}(\beta)\, \sigma_{qg}^{(2)}(\beta) \, .
\label{eq:}
\end{equation}

The meaning of the function $\Sigma_{qg}$ is as follows: when its argument is the highest computed point $\b_{80}$ ($\b_{80}=0.999$ in the case of the function $F_0$), the function $\Sigma_{qg}$ contains the complete contribution to $\sigma_{\rm tot}$ due to the part of the fits (\ref{eq:F1},\ref{eq:F0}) that is beyond the highest computed point, i.e. from the region where our calculation is not derived but extrapolated. As a conservative estimate we take the case of LHC 14 TeV where the partonic fluxes are most enhanced in the high-energy region. We find that $R_{qg}(0.999) \approx 3\times 10^{-5}$ which is completely negligible. In this sense, the uncertainty on the derived constant $c_0$ is of no phenomenological significance for top pair productions at the Tevatron and LHC. However, for applications of Eqs.~(\ref{eq:F1},\ref{eq:F0}) to top pair production at future higher energy hadron colliders or for lighter quark production, like bottom quarks, it would be advisable to re-assess the smallness of the ratio $R_{qg}(0.999)$.

\subsection{Comparison with existing approximations}\label{sec:comparison}

In most past studies of top pair production beyond NLO, the $qg$ reaction has received little attention, and its NNLO correction has, typically, been neglected.  To that end it would be interesting to compare the exact result derived in this paper with the only approximation to $\sigma_{qg}^{(2)}$ derived previously \cite{Langenfeld:2009wd,Moch:2012mk} and implemented in the program Hathor \cite{Aliev:2010zk}. 
\begin{figure}[tbp]
\centering
\includegraphics[width=.95\textwidth]{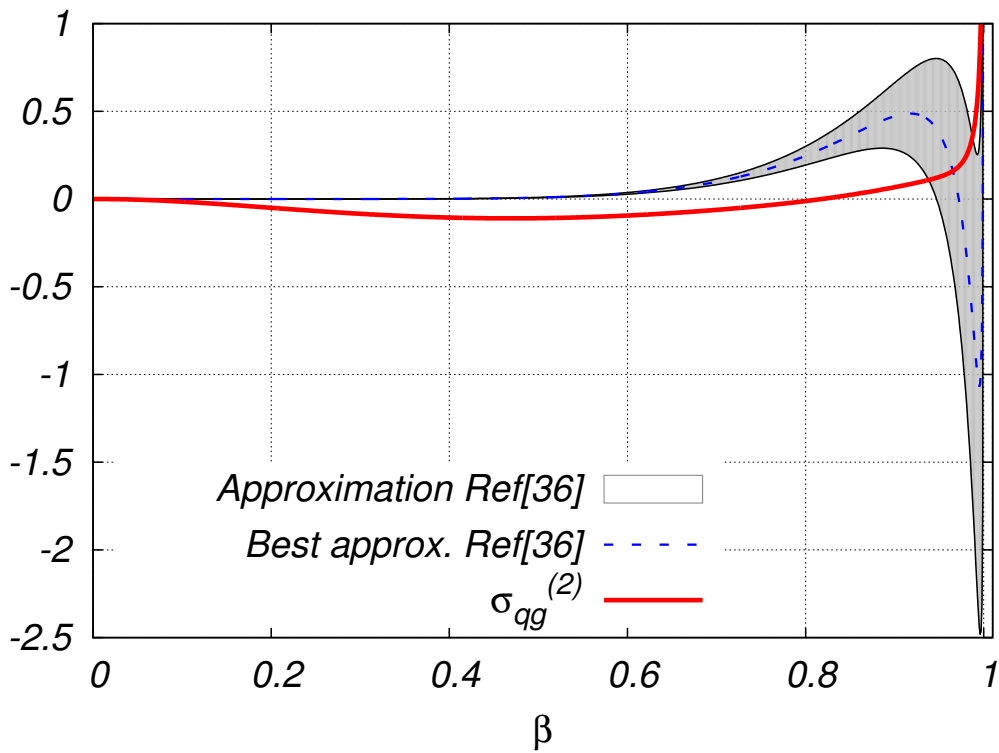}
\caption{\label{comparison-MUV} Comparison of the exact partonic cross-section $\sigma_{qg}^{(2)}$ (solid red) with the approximation of Ref.~\cite{Moch:2012mk} (grey band). The central value (dashed blue) is the ``best" approximation of Ref.~\cite{Moch:2012mk}.}
\end{figure}

In fig.~\ref{comparison-MUV} we plot the envelope of predictions for $\sigma_{qg}^{(2)}$ introduced in Ref.~\cite{Moch:2012mk} (grey band). The prediction of that reference is based on matching the threshold term introduced in Ref.~\cite{Langenfeld:2009wd}
to the high-energy behavior of the cross-section. The spread of the predictions reflects the uncertainty in the prediction of the constant $c_0$ as estimated in Ref.~\cite{Moch:2012mk}. 
\footnote{We remind the reader that both our results and the results in Ref.~\cite{Moch:2012mk} have the same logarithmic behavior in the high-energy limit.}
The blue curve in fig.~\ref{comparison-MUV} denotes the ``best" approximation of Ref.~\cite{Moch:2012mk}. While not explicitly shown in fig.~\ref{comparison-MUV}, the threshold term \cite{Langenfeld:2009wd} is essentially identical to the blue line in the region $\b \lesssim 0.5$. 

It is obvious from fig.~\ref{comparison-MUV} that, except in the limit of extremely large $\b$, the exact result for $\sigma_{qg}^{(2)}$ derived in the present work (red line) has qualitatively different behavior compared to the approximation of Ref.~\cite{Moch:2012mk}. The observed disagreement applies also to the threshold term introduced in Ref.~\cite{Langenfeld:2009wd}. Given the significance of these differences, it is imperative to quantify their phenomenological impact.

In table~\ref{tab:table} we present the contribution to $\sigma_{\rm tot}$ of Ref.~\cite{Moch:2012mk}'s ``best" approximation to $\sigma_{qg}^{(2)}$ (denoted as ``{\it (A+B)/2} " and corresponding to the blue line in fig.~\ref{comparison-MUV}). For its computation we use the program Hathor \cite{Aliev:2010zk} with the same parameters as in the rest of this paper. 

We observe that the difference between the exact result and the approximation of Ref.~\cite{Moch:2012mk} is numerically significant at all collider energies. In particular, at the LHC with c.m. energies of 7 and 8 TeV, the relative difference (with respect to $\sigma_{\rm tot}$) between the approximation of Ref.~\cite{Moch:2012mk} and the exact result can be as large as 5\%. Such a shift in the total hadronic cross-section is very large given that (a) it originates in a subleading channel and (b) it is comparable in size to the total theoretical uncertainty at the LHC. We are therefore led to the conclusion that such large discrepancy is calling into question the usefulness of the high-energy approximation of the heavy flavor production cross-section as a means of describing top pair production at hadron colliders.

\section{Phenomenological predictions}\label{sec:pheno}

Implementing the ${\cal O}(\as^4)$ correction to $\hat\sigma_{qg}$ (\ref{eq:sigma2qg}) in version 1.4 of the program {\tt Top++} we obtain the following ``best" predictions for the Tevatron and LHC 8 TeV:
\begin{eqnarray}
\sigma_{\rm tot}^{\mathrm{NNLO+NNLL}}({\rm Tevatron}) &=& 7.010^{~+0.143\, (2.0\%)}_{~- 0.228\,(3.2\%)}~[{\rm scales}] 
^{~+ 0.186\,(2.7\%)}_{~- 0.122\, (1.7\%)}~[{\rm pdf}]\, , \label{eq:best-res-Tev}\\
&&\nonumber\\
\sigma_{\rm tot}^{\mathrm{(N)NLO+NNLL}}({\rm LHC_{8 TeV}}) &=& 220.4^{~+12.7\, (5.7\%)}_{~- 10.8\,(4.9\%)}~[{\rm scales}] 
^{~+ 5.4\,(2.5\%)}_{~- 5.6\, (2.5\%)}~[{\rm pdf}]\, .~~
\label{eq:best-res-LHC8}
\end{eqnarray}
Theoretical prediction for any other LHC c.m. energy can be easily obtained with version 1.4 of the program {\tt Top++} by adjusting the collider energy in its default LHC setting.

The numbers above are derived in the following way: the partonic reaction $gg\to t\bar t+X$ is included at NLO+NNLL, as in Ref.~\cite{Cacciari:2011hy}. All other partonic channels are now known in full NNLO and are therefore included with the exact NNLO results, including NNLL soft gluon resummation for the $q\bar q\to t\bar t+X$ reaction. We use MSTW2008nnlo68cl pdf set~\cite{Martin:2009iq}, and scale and pdf variations are performed as described in Ref.~\cite{Cacciari:2011hy}.

\begin{figure}[tbp]
\centering
\includegraphics[width=.95\textwidth]{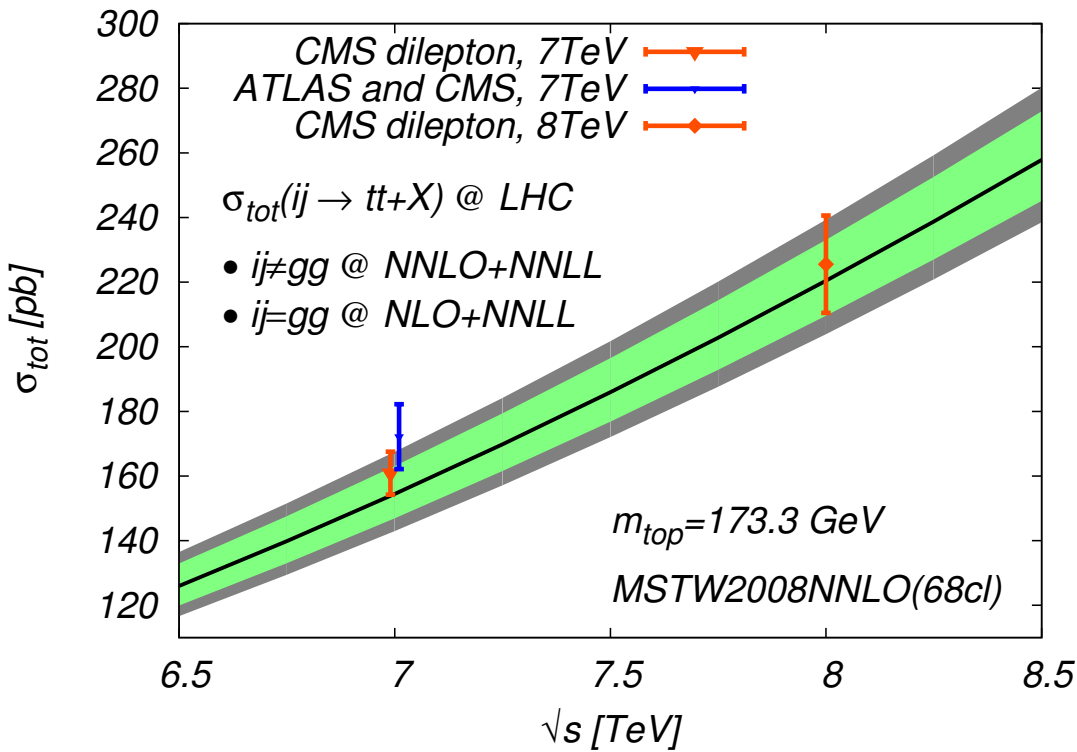}
\caption{\label{LHC-sqrt_s-range} Our ``best" prediction for the LHC as a function of the collider energy. Inner band (light green) represents scale uncertainty; total band is the linear sum of scale and pdf uncertainties. Also shown are the most precise measurements from CMS and ATLAS \cite{:2012bt,CMS:note12-007,ATLAS_and_CMS_x-section2012}.}
\end{figure}

We present our predictions for $m_t=173.3$ GeV. Such value for $m_t$ is consistent with the current best measurements from the Tevatron \cite{Aaltonen:2012ra} ($173.18 \pm 0.94$ GeV), CMS \cite{CMS:note11-018} ($173.36 \pm 0.38 \pm 0.91$ GeV) and with the ATLAS and CMS top mass combination \cite{ATLAS_and_CMS_mass2012} ($173.3\pm 1.4$ GeV). 

The measurements \cite{:2012bt,CMS:note12-007,ATLAS_and_CMS_x-section2012} we compare to, are presented at $m_t=172.5$ GeV (both for 7 and 8 TeV). For a consistent comparison, we translate all measurements to $m_t=173.3$ by rescaling them with a common factor of $0.993512$ \cite{:2012bt}. In principle each measurement should be rescaled with its own scaling factor, however such rescaling is available only for Ref.~\cite{:2012bt}. Given the week dependence of the measurements on the value of the top mass, however, any inconsistency due to this procedure is at the sub-percent level and is thus inconsequential given the size of the experimental and theoretical uncertainties. 

As we discussed in detail in section~\ref{sec:had-lelel-prop}, the inclusion of the NNLO correction to the $qg$ reaction has notable impact on the scale dependence at the LHC. For this reason, with this paper, we update our NLO+NNLL LHC prediction from \cite{Cacciari:2011hy}. In fig.~\ref{LHC-sqrt_s-range} we compare our best prediction (\ref{eq:best-res-LHC8}) with the most precise measurements from CMS and ATLAS \cite{:2012bt,CMS:note12-007,ATLAS_and_CMS_x-section2012}. We note a very good agreement between theory and data at both 7 and 8 TeV. At 8 TeV the total theoretical uncertainty is comparable to the experimental one, while at 7 TeV the experimental uncertainty is almost a factor of two smaller than the total theoretical one, mostly thanks to significantly reduced systematics. We are hopeful that the inclusion of the full NNLO correction in $gg\to t\bar t+X$ in the near future will further reduce the theoretical error.

We also calculate the ratio of the cross-section evaluated at 8 TeV and 7 TeV.  We find that the central value of the ratio, and its uncertainty (evaluated as a restricted scale variation of the ratio, see \cite{Mangano:2012mh}) are not significantly different from the numbers reported in Ref.~\cite{Mangano:2012mh}. 

Before concluding this section we note that the large NNLO correction in $qg\to t\bar t+X$ (relative to the NLO correction in the same reaction) could be indicative of this reaction's possible relevance to the resolution of the $A_{FB}$ puzzle at the Tevatron \cite{Aaltonen:2011kc,Abazov:2011rq}.

\section{Conclusions}

In this paper we calculate the NNLO (i.e. ${\cal O}(\as^4)$) correction to the total top pair production cross-section in the partonic channel $qg\to t\bar t +X$. We follow the computational approach already used in Refs.~\cite{Baernreuther:2012ws,Czakon:2012zr} and compute the partonic cross-section numerically in 80 points on the interval $\beta \in (0,1)$. The numerical precision of the calculation is high, typically below $1\%$. For the practical implementation of the result we have derived analytical fits that have simple analytical form. 

Our result is consistent with its expected endpoint behavior: it vanishes at threshold $\beta=0$ and diverges logarithmically in the high-energy limit $\beta=1$. By imposing the known \cite{Ball:2001pq} exact logarithmic behavior in the high-energy limit we extract the constant in the leading power term. The value of this constant agrees with a recent prediction in Ref.~\cite{Moch:2012mk}. While the uncertainty spreads in each of the two results are not small, the observed agreement is nevertheless an important consistency check on both setups. We have demonstrated that the uncertainty on this constant is completely irrelevant phenomenologically for top pair production at the Tevatron and LHC. 

The phenomenological impact of the NNLO $qg$ correction is moderate. At the Tevatron its only effect is to lower the prediction of Ref.~\cite{Baernreuther:2012ws} with approximately $0.8\%$ which is well within the total theoretical uncertainty. The inclusion of the NNLO $qg$ correction at the LHC lowers the cross-section by approximately $1\%$ while at the same time it decreases the scale uncertainty by about $\pm 2\%$. This is a significant improvement in the theoretical prediction, which agrees well with the most recent LHC measurements at 7 and 8 TeV. 

At present, the dominant source of theoretical uncertainty at the LHC is the lack of the genuinely NNLO correction in the $gg$-initiated reaction. We hope to report results for this last missing at NNLO channel in the near future.

\acknowledgments
We thank S.~Dittmaier for kindly providing us with his code for the evaluation of the one-loop virtual corrections in $qg \to t\bar t q$ \cite{Dittmaier:2007wz}. The work of M.C. was supported by the Heisenberg and by the Gottfried Wilhelm Leibniz programmes of the Deutsche Forschungsgemeinschaft, and by the DFG Sonderforschungsbereich/Transregio 9 ÒComputergest\"utzte Theoretische TeilchenphysikÓ.

\end{document}